# The holographic principle and the Immirzi parameter of loop quantum gravity

M. Sadiq[1], Department of Physics, University of Tabuk, P.O. Box 2072, Tabuk 71451, Saudi Arabia

**Abstract**

The geometrical spectra in loop quantum gravity (LQG) suffer from ambiguity up to the free Immirzi parameter $\gamma$ that is often determined by comparing results from the theory with the established dynamics at the black hole horizon. We address conceptual difficulties associated with such approaches and point out that $\gamma$ can be fixed naively by applying the LQG version of the equipartition rule at a holographic boundary such that the Hawking-Unruh temperature law follows. The value of $\gamma$ derived in this way should possess universal validity. This approach also provides a clue that $\gamma$ could be rooted in the holographic principle.

## A. Introduction

Hawking proved that 'the sum of areas of event horizons never decreases in any classical process involving black holes' [1,2]. This theorem, being analogous to the law of thermodynamic entropy, provided the first clue that black holes can behave thermodynamically. Shortly afterwards, it was shown that black holes obey a set of laws that strongly resemble the laws of thermodynamics [3,4]. Inspired by this similarity, Bekenstein [5] boldly suggested that a black hole has entropy proportional to its horizon area $A$ measured in Planck's units,

$$S = \alpha \frac{A}{G\hbar} \ln 2. \qquad (1)$$

Here $\ln 2$ is the minimal entropy contributed by a bit of information to the horizon and $\alpha$ is the proportionality constant. (We are using units where the speed of light is taken as unity and where temperature is measured in units of energy.) Hawking [6], studying quantum fields in a black hole metric, showed that a black hole horizon should radiate and therefore bear temperature. The emission temperature was calculated as

$$T = \frac{\hbar g_{BH}}{2\pi} = \frac{\hbar}{8\pi GM}, \qquad (2)$$

where $g_{BH} = GMR_S^{-2}$ is the surface gravity at the horizon of a Schwarzschild black hole of mass $M$ and radius $R_S$. Hawking's result further strengthened the thermodynamic description of the black hole horizon. The Hawking temperature formula may be combined with the area-mass relation for a black hole,

$$A = 16\pi (GM)^2, \qquad (3)$$

and the first law of thermodynamics at the horizon,

$$T^{-1} = \frac{\Delta S}{\Delta M}, \qquad (4)$$

to derive the black hole entropy as proportional to the horizon area measured in Planck's units:

---

[1] e-mail: ms.khan@ut.edu.sa

$$S_{BH} = \frac{A}{4G\hbar}. \qquad (5)$$

This formula fixes the constant $\alpha \ln 2$ in Bekenstein's formula exactly at 1/4. Shortly after Hawking's discovery it was realized that his result should not be restricted to black holes alone. Unruh [7] showed that a uniformly accelerating observer through a Minkowski vacuum will perceive a heated horizon with temperature proportional to the observer's acceleration $a$:

$$T = \frac{\hbar a}{2\pi}. \qquad (6)$$

In fact, studies of quantum fields in any spacetime with horizon have credibly shown that all horizons carry temperatures given by the Unruh formula (6) [8-11].

t'Hoopt [12] extended Bekenstein's remarkable insight to a bold conjecture, referred to as the holographic principle. According to this principle, a three dimensional information required to describe an isolated system in a region of space can be represented by the boundary of the region and is limited by the area of this boundary, with the number of microscopic degrees of freedom as finite and proportional to the area of the boundary in Planck's units.

From the Bekenstein-Hawking formula (5) a connection between gravity and thermodynamics can be perceived. Jacobson [13], assuming the relation between area and entropy as universal, showed that Einstein's general relativity has thermodynamical origins. Padmanabhan [14] also showed that gravitational field equations can be derived from thermodynamics. Interestingly, assuming the law of equipartition of energy to hold at a general holographic sphere, Verlinde showed that Newton's gravitational law emerges as an entropic force [15]. Padmanabhan [16] came up with Newton's gravity assuming the law of equipartition of energy and the Hawking-Unruh temperature law to hold at a holographic sphere. Similarly, Smolin [17] derived Newton's law by running the holographic argument together with the equipartition rule within the framework of loop quantum gravity (LQG) [18,19]. It turns out that the Hawking-Unruh temperature is derivable simply from the equipartition of energy applied at a holographic sphere [20]. In this letter, we show that this conceptually simple holographic calculation carried out in the framework of LQG offers a unique way of fixing the Immirzi parameter $\gamma$ of LQG [21] that has been the subject of active investigation since its inception [18,22-27].

The lay out of the paper is the following. Section B briefly reviews some of the hitherto appeared approaches regarding fixing the value of $\gamma$ and discusses some shortcomings resulting therefrom. In section C, a review of Verlinde's derivation of Newton's gravity from the holographic setup is given. The corresponding LQG version due to Smolin is also briefly discussed in the same section. In section D, we turn to the derivation of $\gamma$ in such a way that the LQG version of the equipartition of energy applied at a holographic sphere results in the Hawking-Unruh temperature law. It is in the same section that we will also find a hint that $\gamma$ may be a free parameter linked with the holographic principle. Section E is devoted to results and discussion.

### B. The Immirzi parameter $\gamma$ of LQG

Loop quantum gravity (LQG) is a canonical quantization of the classical gravitational field and uses spin networks as basis for its Hilbert space [18,19]. Spin networks are graphs whose edges carry labels $\{j \in 0, 1/2, 1, ...\}$ as the representations of the gauge group $SU(2)$ of the theory. Amongst various approaches to quantum gravity, LQG seems to be the sole theory that has produced results regarding geometrical spectra from first principle. A key result of LQG is that the area of a given region of space is quantized in such a way that if a surface is punctured by an edge of the spin network, carrying a label $j$, the surface acquires an element of Planck sized area

$$A_j = \beta_j G\hbar. \tag{7}$$

Here $\beta_j = 8\pi\gamma\sqrt{j(j+1)}$ with $\gamma$ as a free undetermined parameter of the theory, called the Immirzi parameter [21]. This parameter is known to have no effect in classical gravity but appears unavoidably in the quantized version of the theory. The actual physical meaning of $\gamma$ is still unknown. The theory however cannot produce useful predictions unless this parameter is worked out. Comparison of the LQG result for the black hole entropy with the known entropy (5) was proposed to provide a means for fixing the Immirzi parameter [28]. In LQG the microscopic states that determine the black hole entropy are the spin network edges puncturing the horizon. Statistically, the dominant contribution to the entropy comes from the lowest possible non-zero spin $j_{min}$ so that the number of edges (with $j = j_{min}$) puncturing the horizon of area $A$ becomes

$$N_{j_{min}} = \frac{A}{\beta_{j_{min}} G\hbar}. \tag{8}$$

Given the multiplicity of the state $j_{min}$ as $(2j_{min}+1)$ the entropy of the black hole is calculated as the logarithm of the dimension of the Hilbert space living on the horizon [23]:

$$S_{LQG} = \frac{A}{\beta_{j_{min}} G\hbar} \ln(2j_{min}+1). \tag{9}$$

For $SU(2)$ as the gauge group of the theory, comparison of (9) for $j_{min} = 1/2$ with the Bekenstein-Hawking relation (5) yields the value of the Immirzi parameter as $\gamma = \ln(2)/\pi\sqrt{3}$ [24,29].

Taking into account the asymptotic quasi-normal mode (QNM) spectrum of a Schwarzschild black hole [30,31] and the Bohr correspondence principle, Dreyer [26] revealed yet another way of fixing $\gamma$. This idea was originally conceived by Hod [32] in the context of black hole dynamics. It was argued that if Bohr's correspondence principle is assumed to be applicable to black holes the radiation or absorption of a quasinormal mode frequency $\omega_{QNM}$ should be consistent with the corresponding change $\Delta M$ in the mass of the black hole, i.e.

$$\Delta M = \hbar\omega_{QNM} = \frac{\hbar\ln 3}{8\pi GM}. \tag{10}$$

From relations (3), (7), and (10), one is able to fix $\gamma$ as

$$\gamma = \frac{\ln(3)}{2\pi\sqrt{j_{min}(j_{min}+1)}}. \tag{11}$$

However, to comply with the black hole entropy formula (5), it became necessary to consider the edges with $j_{min} = 1$ as dominant, thereby fixing $\gamma$ at $\ln(3)/2\pi\sqrt{2}$. But the dominant contribution as coming from the $j = 1$ transitions in the $SU(2)$ framework is not one would expect statistically. Dreyer therefore proposed that instead of $SU(2)$ one should adopt $SO(3)$ as the gauge group of LQG.

Thereafter, attempts were made to save $SU(2)$ as the relevant gauge group and to formulate convincing explanation about why edges with $j = 1$ rather than $j = 1/2$ contribute dominantly to the black hole entropy [25,33,34]. Corichi, for instance, argued that altering the group of the theory from $SU(2)$ to $SO(3)$ would not be that appealing of an idea if fermions were to be accomudated in the theory [25]. Retaining *SU*(2) as the relevant working group and invoking the local fermion number conservation, Corichi explained why the $j=1$ processes contribute dominantly. Whereas the value of $\gamma$ for $j=1$ processes in the *SU*(2) framework should have been different, Corichi reported the same value as obtained by Dreyer. Mimicking a QNM frequency as *SU*(2) oscillating system [27] we modified the value of $\gamma$ to be twice as much as the value reported by Dreyer and Corichi, where the extra factor of 2 is reminiscent of the fact that *SU*(2) is the double map of *SO*(3).

We conclude this section by mentioning few issues that still remain unexplained. The problems are the following: (i) As outlined earlier, the semi-classical entropy formula (5) was obtained by incorporating the classical area-mass relation (3), the first law of thermodynamic (4) at the boundary, and the Hawking temperature formula (2) that by itself does not include quantum properties of gravity. On the contrary, the LQG's quantum description of the black hole entropy (9) appears to inherit no such low energy attributes of a black hole. Therefore, making exact correspondence of the quantized description in (9) with the semi-classical result (5) is questionable. (ii) The approaches discussed above leave us with the ambiguity as to which group, *SU*(2) or *SO*(3), should be adopted as the gauge group of the theory. (iii) The Immirzi parameter $\gamma$ has not been fixed in an environment other than that of black hole, whereas the universality of gravity suggests that $\gamma$ should be fixed in such a way as to possess universal validity. (iv) Being content only with fixing the value of $\gamma$, none of the approaches mentioned above explain why $\gamma$ is absent in classical gravity but appears unavoidably in the quantized version of the theory.

### C. Holography and Newton's Gravity

Verlinde [15] conjectured that Newton's gravity can be explained as an entropic force that arises due to changes in information associated with the positions of material objects near a holographic sphere. Following Bekenstein's seminal idea [5] regarding the black hole entropy, Verlinde assumed that when a test particle of mass $m$ approaches at a distance of one Compton wavelength $\Delta x = \hbar/m$ near a holographic two-sphere from the side where classical spacetime has just emerged, the entropy of the sphere will change by an amount $2\pi$, which one can as

$$\Delta S = 2\pi \frac{m}{\hbar} \Delta x \tag{12}$$

to exhibit proportionality of the change in entropy to the mass added. In order for the screen to have entropy there must be a non-zero temperature $T$ associated with the screen. The change in entropy, and hence the change in energy $\Delta U$, means that there is a force $F$ acting on the particle that can be deduced from the first law of thermodynamics

$$\Delta U = F \Delta x = T \Delta S. \tag{13}$$

This effective force is called the entropic force that arises in a system with large degrees of freedom by the statistical tendency to increase its entropy. The entropic force requires no intermediate fundamental field for its mediation. Verlinde further assumed that the energy $E$ of the system is divided evenly over $N$ degrees of freedom (bits of information) residing on the screen. If one writes $E = M$, where $M$ is the mass that would emerge in the part of space bounded by the screen, then the equipartition rule determines the temperature of the screen as

$$T = \frac{2M}{N}. \tag{14}$$

One substitutes (12) and (14), with $N = 4\pi R^2/G\hbar$, in (13) to determine the entropic force, that turns out to be the familiar Newton's law of gravity

$$F = \frac{GMm}{R^2}. \tag{15}$$

Following the LQG version of Verlind's holographic set of equations, Smolin [17] obtained the gravitational law in the form

$$F = \frac{GM}{R^2}\left(\frac{\hbar}{\Delta x}\frac{2(\ln 2)^2}{\pi}\right). \tag{16}$$

In this derivation Smolin considered $j_{\min} = 1/2$ edges as dominant and used the value of the Immirzi parameter as $\gamma = \ln(2)/\pi\sqrt{3}$, which was established only in the black hole environment. Smolin interpreted the quantity in the brackets, that envelops the dimensionless fudge factor $f = 2(\ln 2)^2/\pi$ also, as representing the passive gravitational mass $m$ of a particle near the holographic screen within its Compton wavelength approximately equal to $\Delta x$. Smolin maintained that the fudge factor $f$ was important because the Compton wavelength only gave an approximation as to how high must the particle be above the surface in order to be considered as exterior to the surface.

### D. Examining $\gamma$ from the holographic perspective

For a comprehensive discussion regarding connection between the holographic formulation of quantum gravity and loop quantum gravity we refer to Smolin's work [17,35]. Following [17], we show that the derivation of Hawking-Unruh temperature from the LQG version of the holographic setup offers a radically different answer to the Immirzi parameter question.

Let us consider a general holographic two-sphere of radius $R$ and area $A$ that encloses a spherically symmetric distribution of mass $M$. Holography requires that any three dimensional information enclosed by a holographic sphere is to be considered as residing on the surface. As noted earlier, in LQG the number of microstates on the boundary can be determined from the number of edges $N_{j_{\min}}$ puncturing the surface, where the boundary can be assumed to be dominated by the lowest possible spins $j_{\min}$ if entropy is to be maximized. Any exchange of energy $\Delta M$ at the surface can be viewed as attachment or detachment of an edge $j_{\min}$. Since a change in energy of the boundary is associated with a change in its entropy, there must be a finite temperature $T$ associated with the boundary. Moreover, for spherical symmetry one can assume that the energy $E = M$ of the screen is distributed evenly over the edges. This allows us to implement a version of the equipartition rule on edges at the screen:

$$M = \frac{1}{2}N_{j_{\min}}T, \tag{17}$$

Substituting $N_{j_{\min}}$ from (8), with $A = 4\pi R^2$, into (17) one obtains temperature of the screen as

$$T = \beta_{j_{\min}}\frac{\hbar g}{2\pi}. \tag{18}$$

Here $g = GMR^{-2}$ is the surface gravity at the holographic two-sphere. Equation (18) is precisely the Hawking-Unruh temperature, a local accelerated observer would experience at the boundary, provided one choses $\beta_{j_{\min}}$ as unity. Obviously, this means that the holographic principle yields the correct temperature law at the boundary provided the fundamental quantum of area is chosen exactly as the Planck area. The value of $\gamma$ must therefore be chosen as to yield $\beta_{j_{\min}} = 1$:

$$\gamma = \frac{1}{8\pi\sqrt{j_{\min}(j_{\min}+1)}}. \tag{19}$$

Indeed, this value of $\gamma$ is true at any holographic sphere, including the black hole horizon, and therefore possesses universal validity. Assuming $SU(2)$ as the underlying group (that has to be fixed from elsewhere) and taking the spin-1/2 edges as dominant, the value of $\gamma$ can be fixed as $1/4\pi\sqrt{3}$. We remark that with this new value of $\gamma$, Smolin's result (16) will be modified to

$$F = \frac{GM}{R^2}\left(\frac{\hbar}{\Delta x}\frac{\ln 2}{2\pi}\right), \tag{20}$$

with a new fudge factor $f = \ln 2/2\pi$. One would notice that, instead of $2\pi$, had Verlinde considered the minimal entropy contributed by a bit to be $\ln 2$, he would have arrived at the same result as (20).

We now intend to explore a peculiar aspect of the holographic principle with regard to the derivation of Newton's gravity. Recall that Verlinde took the minimal area element to be exactly the Planck area in his derivation. Let $S$ be the entropy of the holographic sphere when its area is measured in Planck units. But one can easily prove that the holographic principle allows for Newton's law to follow even when the fundamental area element is ambiguous up to a free unknown parameter. To see how this happens, let us suppose we have a microscopic theory of geometry (for instance, LQG) in which the minimal quantum of area is given by $\beta G\hbar$, with $\beta$ as an arbitrary scale factor. The number of bits on the holographic sphere of a given area $A$ should now become $N_\beta = A/\beta G\hbar$. Consequently, every other quantity dependent on this number should be modified. Since entropy is proportional to the number of bits, changing this number on the screen would require entropy of the screen to be rescaled as $S_\beta = S/\beta$. Similarly, the entropic force will be read off as $F_\beta = T_\beta \Delta S_\beta / \Delta x$. It immediately follows that the scale factor $\beta$ drops out and Newton's law is restored, i.e., $F_\beta = F = GMm/R^2$. This affirms that the actual operational quantum of area is the Planck area and that any ambiguity present at the quantum level does not affect the classical result. Likewise, the same argument could be repeated for the LQG result (20); one would just need to replace $\beta$ by $\beta_j$.

The outcome of the foregoing discussion is that the holographic principle recognizes the bare Planck area as generic to the physical laws, and allows for arbitrariness at the microscopic scale such that the physical results it leads to are independent of any arbitrary parameter. Since holography is naturally linked to LQG [17,35], one may wonder whether the appearance of a similar ambiguity in the geometric spectra of LQG is merely a coincidence or is it a holographic feature that has made its way into the quanta of geometry to exhibit its relevance to holography.

### E. Discussion

The equipartition of energy imposed on the edges of LQG puncturing a general holographic sphere reproduces the Hawking-Unruh temperature provided the physically relevant area element is chosen precisely as the Planck area, which in turn fixes the Immirzi parameter $\gamma$ at a value distinct from what has been reported earlier. Fixing the value $\gamma$ in this way permits it to be ubiquitously valid at every holographic sphere, including the black hole horizon, and whereby no ambiguity arises over the group structure of the theory. Further, in the holographic approach, it is again the Planck area that determines Newton's gravity and, more significantly, any ambiguity present at the quantum level turns away in (at least non-relativistic) classical gravity—a characteristic that is reminiscent of $\gamma$. Since holography is tied up with LQG, it seems worth delving deep to explore whether $\gamma$ could indeed be identified as a parameter connected to holography. If found so from another match, the case for LQG to be a successful candidate theory of quantum gravity would even be strengthened.

Recall that, apart from the Hawking temperature (2) and the first law (4), the area-mass relation (3) is crucial to the derivation of the black hole entropy formula (5). Such a relationship however is lacking in the case of a holographic sphere other than the black hole horizon. The absence of such a relationship prohibits us to formulate a unified scheme explaining both temperature and entropy of a general holographic sphere. Thus, even though the Hawking-Unruh temperature law is the same for every holographic sphere it is not clear what entropy can be attributed to a general holographic sphere.

Viewed from the holographic perspective, there exist now reasons to revise the formal LQG description of the black hole entropy given by (9). First, in light of the newly established value of $\gamma$, one notices that the LQG result (9) actually describes rescaled entropy akin to $S_\beta$ and, therefore, cannot be equal to entropy $S$ measured with the bare Planck units—note that embodied in the Bekenstein-Hawking formula (5) is the bare Planck area because it is inseparable from the Hawking temperature formula (2). Second, if entropy was merely 'equal' to the logarithmic measure of the dimensionality of the Hilbert space residing on a boundary then there would essentially be no difference between the entropy of a general holographic sphere and that of the black hole horizon. Obviously, this is not the result one would expect. We therefore propose that, with bare Planck units, the entropy of a general holographic sphere (including the black hole horizon) should not be evaluated as 'equal' but 'proportional' to the logarithm of the number of dimensions of the Hilbert space living on the boundary. In this way the LQG calculation for a general holographic sphere will effectively coincide with the Bekenstein formula (1). However, the proportionality constant cannot be fixed unless one knows, besides (4) and (6), an area-mass relation like (3) for a general horizon. Fortunately, for the black hole horizon, the constant uniquely turns out to be such that the entropy is one-quarter the horizon area in Planck units.

## Acknowledgement

I acknowledge the financial support for this project from the Deanship of Scientific Research (DSR), University of Tabuk, Tabuk, Saudi Arabia, under grant no. S-0026-1436. I also express my sincere thanks to Dr. Taymour A. Hamdallah and Dr. A. Abdesselam for their help in pursuing this project.